\documentclass{jpp}

\usepackage{graphicx}%
\usepackage{dcolumn}%
\usepackage{bm}%

\usepackage[utf8]{inputenc}
\usepackage[T1]{fontenc}
\usepackage{mathptmx}
\usepackage{siunitx}

\usepackage{tikz}
\usepackage{color}
\usepackage[]{xcolor}
\usepackage{amsmath}

\shorttitle{Turbulence and transport in LAPD mirrors}
\shortauthor{Phil Travis and Troy Carter}

\title{Turbulence and transport in mirror geometries in the Large Plasma Device}

\author{Phil Travis\aff{1}
  \corresp{\email{phil@physics.ucla.edu}},
  \and Troy Carter\aff{1}\aff{2}}

\affiliation{\aff{1}Department of Physics and Astronomy, University of California, Los Angeles, California 90095-1547
\aff{2}Oak Ridge National Laboratory, Oak Ridge, TN 37830}

\begin{document}

\maketitle

\begin{abstract}
Thanks to advances in plasma science and enabling technology, mirror machines are being reconsidered for fusion power plants and as possible fusion volumetric neutron sources. However cross-field transport and turbulence in mirrors remains relatively understudied compared to toroidal devices. Turbulence and transport in mirror configurations were studied utilizing the flexible magnetic geometry of the Large Plasma Device (LAPD). Multiple mirror ratios from $M=1$ to $M=2.68$ and three mirror-cell lengths from $L=3.51$m to $L=10.86$m were examined.
Langmuir and magnetic probes were used to measure profiles of density, temperature, potential, and magnetic field. The fluctuation-driven $\tilde{E} \times B$ particle flux was calculated from these quantities. Two probe correlation techniques were used to infer wavenumbers and two-dimensional structure.
Cross-field particle flux and density fluctuation power decreased with increased mirror ratio.
Core density and temperatures remain similar with mirror ratio, but radial line-integrated density increased.
The physical expansion of the plasma in the mirror cell by using a higher field in the source region may have led to reduced density fluctuation power through the increased gradient scale length. This increased scale length reduced the growth rate and saturation level of rotational interchange and drift-like instabilities.
Despite the introduction of magnetic curvature, no evidence of mirror driven instabilities — interchange, velocity space, or otherwise — were observed. For curvature-induced interchange, many possible stabilization mechanisms were present, suppressing the visibility of the instability.
\end{abstract}

\section{\label{sec:intro}Introduction}

Historically, mirror research has prioritized the main issues with mirror confinement: stabilizing the interchange instability, stabilizing velocity-space (loss-cone-driven) instabilities, and minimizing axial electron heat losses. Nevertheless cross-field transport remains an important topic in magnetic-confinement fusion reactor development, in both linear and toroidal geometries. Insight into edge-relevant turbulence, and its coupling to interchange and other mirror-driven instabilities, performed in a basic plasma science device may be useful for a mirror-based reactor. Although not at fusion-relevant core temperatures or densities, the Large Plasma Device (LAPD) operates at conditions similar to the edge of fusion devices and can provide insight into the physical processes in that region. Mirror machines are once again rising in prominence as a candidate for commercial fusion reactors with the advent of highly-funded commercial ventures and high-field high-temperature superconducting magnets \citep{WHAM, BEAM}, so development of a functional understanding of cross-field transport in mirrors is imperative. A characterization of edge fluctuations has been undertaken, with emphasis on interpreting these fluctuations within the context of mirror. 

Non-classical cross-field particle transport is often caused by low-frequency, large-amplitude fluctuations. These fluctuations are the result of various instabilities. One such process is the "universal" drift instability, which appears in the presence of a density gradient and finite resistivity. Drift wave turbulence and the effect on transport has been extensively studied in the past \citep{Horton_1999, Tynan_review_2009}. In the presence of sufficiently high rotation or sheared flow, rotational interchange and the Kelvin-Helmholtz instabilities also contribute or couple to these fluctuations. 

Various gradient-, rotation-, and shear-driven instabilities (and suppression of such) have been studied previously in the LAPD experimentally \citep{Schaffner_2012, Schaffner_turbulence_2013, Schaffner_2013} and in simulations using BOUT, a 3d fluid turbulence code, and an eigenvalue solver \citep{Popovich_2010}. The LAPD has a sufficiently high spontaneous rotation rate that rotation-driven instabilities may be excited without artificial drive. Simulations using BOUT++ \citep{Friedman_2013} have also suggested that a rapidly growing nonlinear instability may dominate over all other linear instabilities.

Imposing a magnetic mirror configuration introduces magnetic curvature. The alignment of the curvature vector with a pressure gradient vector component causes the flute-like interchange instability if no stabilization mechanism is present. This interchange mode could couple to finite $k_\parallel$ drift waves. The coupling of drift waves to curvature-induced interchange modes has been studied in toroidal devices such as TORPEX \citep{Poli_experimental_2006, Fasoli_electrostatic_2006}, where curvature was seen as the driving component for the unstable drift-interchange modes. Drift-like fluctuations have also been observed in the GAMMA-10 mirror \citep{Mase_1991, Yoshikawa_2010}. Flute-like modes and drift waves have been studied in other linear devices, such as Mirabelle \citep{Brochard_transition_2005}, where the appearance of flute-like modes or drift waves were controlled by varying the field and limiter diameter. 

The rotational interchange and curvature interchange can both be flute-like modes. Rotational interchange (also called the ``centrifugal instability") is driven by the aligned centrifugal force and pressure gradient vectors, but curvature-driven interchange is instead driven by magnetic curvature and is typically referred to as simply the ``flute" or ``interchange" instability. Rotational interchange \citep{Jassby_transverse_1972} has been observed in the LAPD in the past \citep{Schaffner_2012, Schaffner_2013}, and the curvature-driven interchange instability has been observed in many other mirror machines \citep{wickham_curvature-induced_1982, ferron_interchange_1983, Post_1987}.

Biasing or modifying the electrical connection of the plasma with the end wall has proven to be a important actuator in many mirror machines such as TMX-U \citep{Hooper_1984}, GAMMA-10 \citep{Mase_1991}, and GDT \citep{Bagryansky_2003, Bagryansky_2007, Beklemishev_2010}, and will be utilized on WHAM \citep{WHAM}. Active biasing was not attempted in this study, but the intrinsic rotation and strong electrical connection to the source region may provide a useful analog for edge biasing in other mirror machines.

The LAPD exhibits a high degree of turbulence so it is difficult to identify the dispersion relation of the modes that are present. Nevertheless, the LAPD has good coverage of perpendicular spectra using correlation-plane techniques, and some measure of parallel spectra using the correlation between two axially-separated probes. A space-time spectral characterization of the many instabilities present in this low beta, moderate aspect ratio, gas-dynamic trap regime is attempted.

This goal of this study was to investigate the changes to turbulence and transport in LAPD mirror configurations. Of particular interest were the potential coupling of the interchange instability with drift waves or other modes, and the effect of the mirror geometry on cross-field particle flux. Presented is a characterization of the observed modes and the effect of introducing a mirror geometry.
This paper is organized as follows. Sec. \ref{sec:methods} discusses the configuration of the LAPD and the diagnostics used. Sec. \ref{sec:changes} covers the changes seen when imposing a magnetic mirror configuration on profiles, particle flux, drift waves, turbulence, and magnetic fluctuations. Sec. \ref{sec:structure} explores the changes in 2d (x-y plane) structure. Sec. \ref{sec:discussion} discusses the active and expected instabilities and reasons for their modification. Sec. \ref{sec:conclusion} summarizes the study and discusses the requirements for a deeper investigation.

\section{\label{sec:methods}Device configuration}

\subsection{\label{sec:sub_exp}The Large Plasma Device (LAPD)}
The Large Plasma Device (LAPD) is a 20 meter long, 1 meter diameter basic plasma device at UCLA \citep{LAPD_BaO}. The LAPD has a variable magnetic field, from 250G to 1.6 kG and can be varied axially. Probes inserted into the plasma can collect high-resolution, temporal information on density, temperature, potential, and magnetic field fluctuations. In this study, the plasma was formed using an emissive, $72$ cm  diameter barium-oxide (BaO) cathode (mapped to 60 cm in a flat field) and a $72$ cm diameter, 50\% transparent molybdenum anode that accelerate electrons across a configurable $40-70$V potential; voltages of $60$ and $63$V were used in this study. The source has since been upgraded to a lanthanum hexaboride (LaB6) cathode \citep{LAPD_LaB6} that enables access to higher-density, higher-temperature regimes, but all the data in this study are from plasma formed by a BaO cathode. 

\begin{figure*}
    \centering
    \includegraphics[width=360pt]{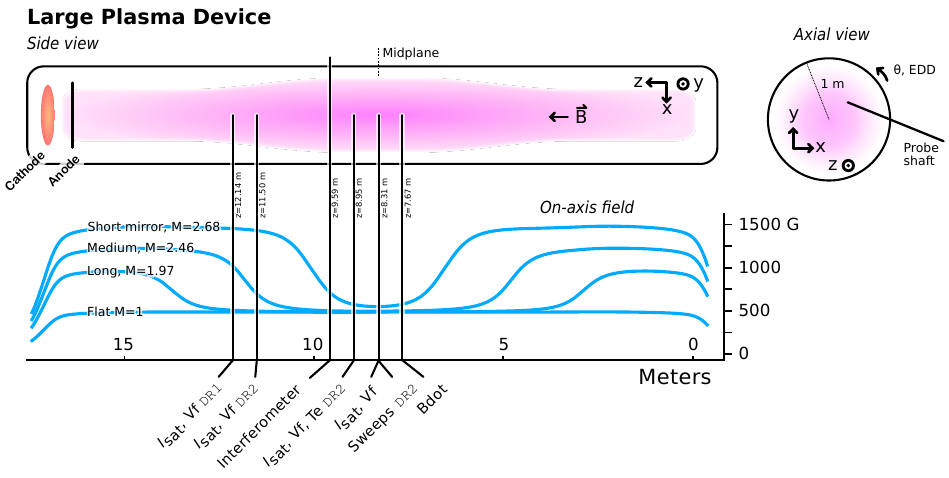}
    \caption{Cartoon of the Large Plasma Device and the coordinate system used. Only four of the eleven mirror configurations studied are plotted for clarity (mirrors of the same length have similar shapes and simply scale with mirror ratio). Diagnostic set varied by datarun; unlabeled diagnostics were used in both dataruns.}
    \label{fig:magnetic_geometry}
\end{figure*}

The flexible magnetic geometry of the LAPD was used to construct a variety of magnetic mirror configurations. The discharge current, fill pressure, and other machine parameters were held constant. The typical plasma parameters observed in this study can be seen in table \ref{tab:plasma-parameters}. 
Data in several mirror ratios and lengths were collected (see table \ref{tab:fields}) but emphasis is placed on the short cell because the highest mirror ratio possible ($M=2.68$) with a $500$ Gauss midplane field could be accessed and probes were able to be placed outside of the mirror cell. An overview of the axial magnetic field for the the short mirror configurations and probe locations can be seen in fig. \ref{fig:magnetic_geometry}. 2- or 3-cell mirror configurations were also explored but are not examined in this study. All results presented below are from the short mirror cell configuration unless otherwise specified.

\begin{table}
 \centering
 \begin{tabular}{l l l l l}
 Mirror length & \multicolumn{4}{l}{Mirror ratios ($M$)} \\
 \hline
 Flat & 1 & & & \\
  3.51 m (short) & 1.47 & 1.90 & 2.30 & 2.68 \\ 
 7.03 m (medium) & 1.49 & 1.98 & 2.46 & \\
 10.86 m (long) & 1.47 & 1.97 & 2.44 & \\
 \end{tabular}
\caption{\label{tab:fields}Magnetic mirror lengths and ratios. The lengths are measured where the curvature changes sign and the ratio is the maximum divided by the minimum. Approximately $3.5$m must be added to the length if the good-curvature region is included. In the case of small asymmetries, the field strengths were averaged before calculation of the mirror ratio.}
\end{table}

\subsection{\label{sec:sub_diagnostics}Diagnostics}

\begin{figure}
    \centering
    \includegraphics[width=350pt]{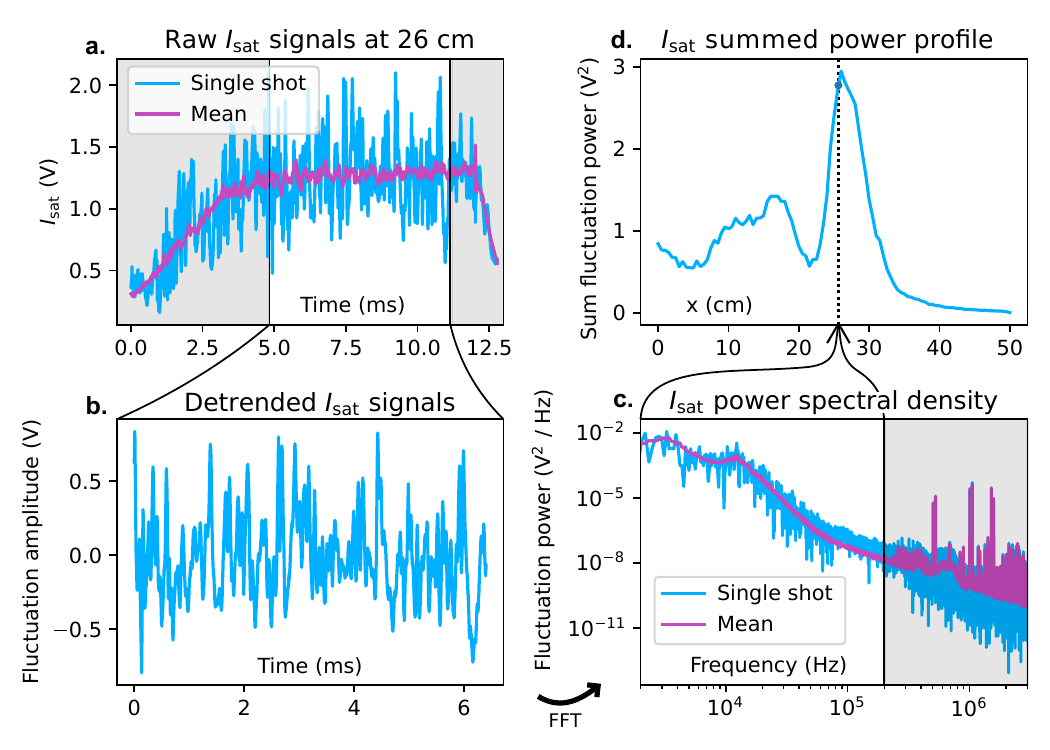}
    \caption{Raw data and basic processing steps for LAPD probe diagnostics as demonstrated by an $I_\text{sat}$ trace from a \texttt{DR1}, $M=1$ mirror at 26 cm. Data are truncated from 4.8 to 11.2 ms (a) and detrended (b). Power spectral density is calculated (c), and a power profiles can be constructed (d). The shaded regions are excluded from this analysis.}
    \label{fig:raw-signals-plots}
\end{figure}

All diagnostics were recorded with a effective sampling rate of 6.25 MHz (16-sample average at 100 MSPS) and a spatial resolution of 0.5 cm. When necessary, averaging over time is done in the approximate steady-state period of the plasma discharge ($4.8$ to $11.2$ ms from the $1$ kA trigger signal). Unless otherwise noted, all data presented will be from probes inside the mirror region ($z \approx 7$m). An example of a raw $I_\text{sat}$ signal and processing steps can be seen in fig. \ref{fig:raw-signals-plots}. The raw signals are detrended by subtracting the mean across shots to obtain the fluctuations only. FFTs are then taken of these fluctuations for calculating power spectra and cross-correlated quantities. Frequencies above 200 kHz are dominated by electronics and instrumentation noise and thus are also ignored. Fluctuation power profiles can then be constructed.

The data presented were collected in two phases. The first phase ("datarun"), \texttt{DR1}, collected Langmuir probe ($I_\text{sat}$ and Vf) and magnetic fluctuation ("Bdot") \citep{Everson_design_2009} traces. 50 shots were taken at each position for every configuration. 
The second phase, \texttt{DR2}, was conducted with a similar set of diagnostics focused on temperature measurements (swept and triple probe) and 2d x-y structure. 15 shots were taken at each position, except for Langmuir sweeps with 64 shots. When appropriate, all data for each position were shot-averaged.
"$I_\text{sat}$" will be used interchangeably with "density" and be presented with units of density (assuming a flat $T_e = 4.5$ eV profile).

\begin{table}
    \centering
    \begin{tabular}{l l l l l}
        Cathode radius (M=1) & $x_c$ & $30$ && cm \\
        Machine radius & $R$ & $50$ && cm \\
        Plasma length & $L$ & $\sim 17$ && m \\
        Primary species && He-4 1+ \\ 
        Electron-helium mass ratio && $1.37 \times 10^{-4}$ \\
        Neutral pressure && $6 - 20 \times 10^{-5}$ && Torr \\
        \hline
        Quantity &  & Core & $x=x_\text{PF}$ & Unit \\
        \hline
        Density & $n_e$ &  $1.25 \times 10^{12}$ & $ 0.6 \times 10^{12}$ & $\text{cm}^{-3}$ \\
        Ion temperature & $T_i$ & $\sim 1$ & — & eV \\
        Electron temperature & $T_e$ & $4$ & $5$ & eV \\
        Beta (total) & $\beta$ & $9 \times 10^{-4}$ & $6 \times 10^{-4}$ & \\
        Midplane magnetic field & $B_\text{mid}$ & $500$ & — & G \\
        Plasma freq & $\Omega_{pe}$ & $10$ & $7.1 $& GHz \\
        Ion cyclotron freq & $\Omega_{ci}$ & $200$ & — & kHz \\
        Electron cyclotron freq & $\Omega_{ce}$ & $1.4$ & — & GHz \\
        Debye length & $\lambda_D$ & $0.013$ & $0.021$ & mm \\
        Electron skin depth & $\lambda_{e}$ & $30$ & $43$ & mm\\
        Ion gyroradius & $\lambda_{ci}$ & $5.8$ & — & mm \\
        Electron gyroradius & $\lambda_{ce}$ & $0.13$ & 0.15 & mm \\
        Ion thermal velocity & $\bar{v}_i$ & $6.94$ & — & km/s\\
        Electron thermal velocity & $\bar{v}_{e}$ & $1190$ & $1330$ & km/s \\
        Sound speed & $c_s$ & $13.0$ & $13.9$ & km/s\\
        Alfvén speed & $v_a$ & $446 - 1140$ & $ - 1620$ & km/s \\
        Ion sound radius & $\rho_s$ & $65$ & $69$ & mm\\
        Ion-ion collision freq & $\nu_{ii}$ & $730$ & $380$ & kHz \\
        Electron-ion collision freq & $\nu_{ei}$ & $6.77$ & $2.59$ & MHz \\
        Electron collision freq & $\nu_{ee}$ & $9.57$ & $3.66$ & MHz \\
        Ion mean free path & $\lambda_{i,\text{mfp}}$ & $26$ & $50$ & mm \\
        Electron mean free path & $\lambda_{e, \text{mfp}}$ & $175$ & $512$ & mm \\
        Spitzer resitivity &  $\eta$ & $192$ & $146$ & \SI{}{\micro \ohm \meter} \\
        
    \end{tabular}
    \caption{LAPD machine information and plasma parameters in the core and peak-fluctuation region ($x=x_\text{PF}$) at the midplane in this study. Dashed quantities are assumed to be identical to core quantities.}
    \label{tab:plasma-parameters}
\end{table}

\section{\label{sec:changes}Mirror-induced changes}

\subsection{Profile modification}

\begin{table}
    \centering
    \begin{tabular}{c c c c c c}
         Mirror ratio & 1 & 1.47 & 1.90 & 2.30 & 2.68 \\
         \hline
         Scale factor & 1 & 1.21 & 1.38 & 1.52 & 1.64\\
         $x_{c}$ (cm) & 30 & 36 & 41 & 45 & 49 \\
         $x_{PF}$ (cm) & 26 & 32 & 36 & 40 & 43 \\
    \end{tabular}
    \caption{$x_c$ and $x_{PF}$ locations for each mirror ratio when scaled by the expected magnetic expansion.}
    \label{tab:x_PF}
\end{table}

\begin{figure}
    \centering
    \includegraphics[width=200pt]{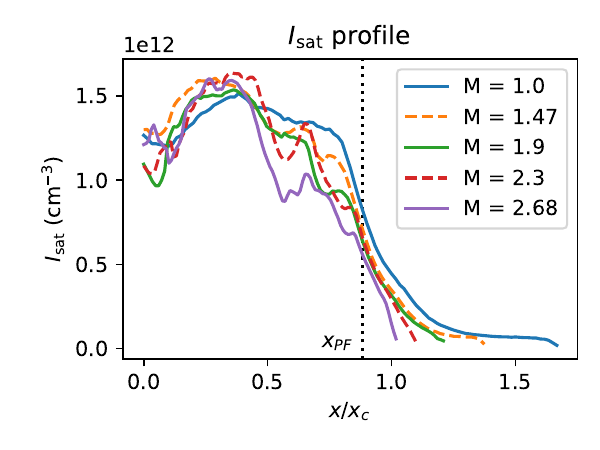}
    \caption{Midplane $I_\text{sat}$ profile, shot-averaged and time-averaged from 4.8 to 11.2 ms (assumed of $T_e = 4.5$ eV based on triple probe and Langmuir sweep measurements). Effective area was calibrated using a nearby interferometer. Profile shape remains similar in the core and gradient region when mapped to the cathode radius $x_c$. The dips in profiles at higher $M$ below $x=x_\text{PF}$ are of unknown origin and are not the focus of this study. Shot-to-shot variation is less than 5\% for $x \leq 0.95 x_c$ and less than 9\% for $x \leq 1.4 x_c$ for all cases.}
    \label{fig:isat_profile}
\end{figure}

\begin{figure}
    \centering
    \includegraphics[width=200pt]{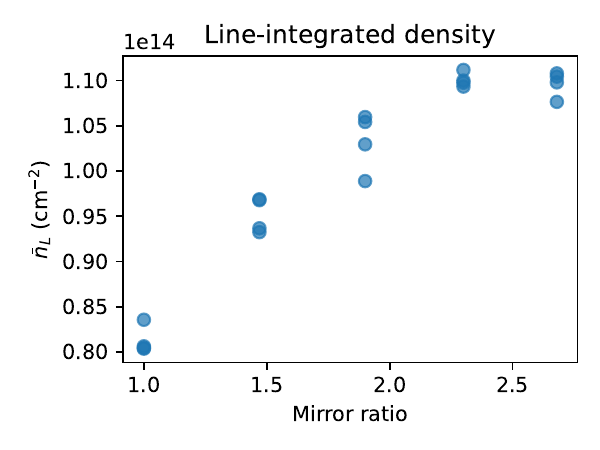}
    \caption{Line-integrated density as measured by a 56 GHz heterodyne interferometer as a function of mirror ratio, taken from four discharges for each mirror configuration. Density increases up to a mirror ratio of 2.3 where it appears to level off. The interferometer is located in the mirror cell bad-curvature region at 9.59m, 1.3m closer to the cathode from the midplane.}
    \label{fig:density-line}
\end{figure}

\begin{figure}
    \centering
    \includegraphics[width=200pt]{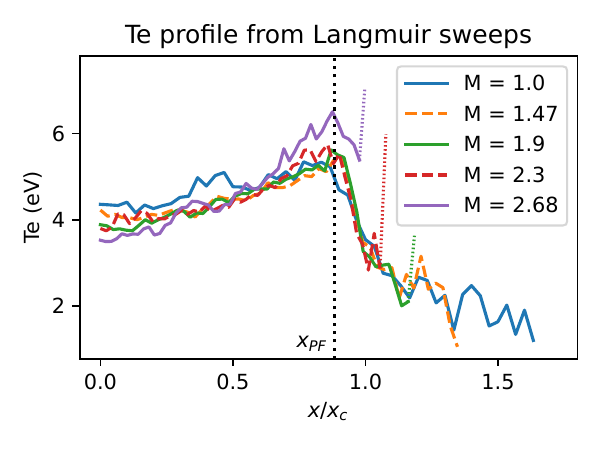}
    \caption{$T_e$ from Langmuir sweeps (\texttt{DR2}) at the midplane. Triple probe results are nearly identical. The increased temperatures directly at the plasma edge, indicated by dotted portions of the curves, are likely artifacts caused by sheath expansion in lower densities. Changes in mirror ratio lead to at most 25\% change in $T_e$. The plasma is collisional and isotropic.}
    \label{fig:Te-sweeps}
\end{figure}

\begin{figure}
    \centering
    \includegraphics[width=200pt]{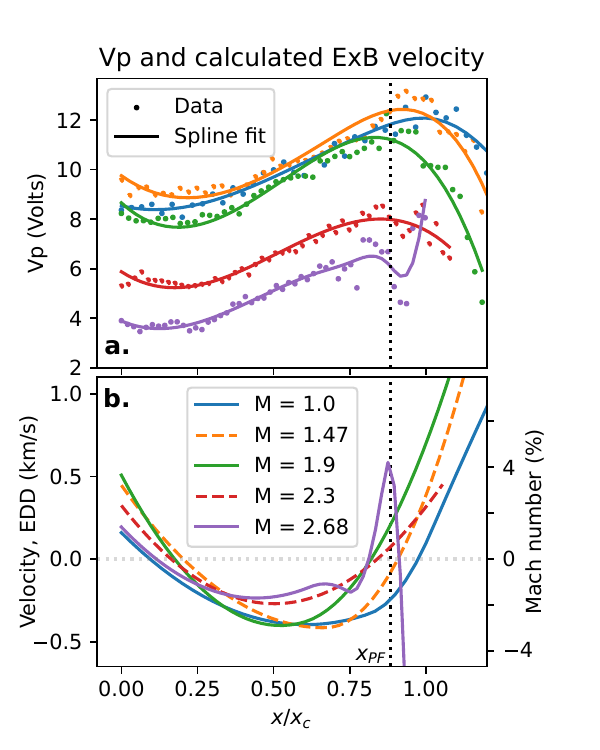}
    \caption{Plasma potential (a) and derived $\boldsymbol{E \times B}$ velocity profiles (b) from Langmuir sweeps at the midplane. $x/x_c > 1.2$ has been excluded from the graph for greater clarity in the core and gradient region. The electric field was calculated by taking the gradient of the spline-smoothed plasma potential profile. The Mach number (in percent) is calculated using the approximate sound speed evaluated at $x=x_{PF}$ (tab. \ref{tab:plasma-parameters}). The overall structure of the flows does not appreciably change when mirror ratio is varied.}
    \label{fig:Vp_ExB}
\end{figure}

Because the field at the plasma source increases with $M$, the midplane plasma expands by a factor of $\sqrt{M}$. The physical locations of the peak fluctuation region -- $x_{PF}$ (maximum gradient) -- and the cathode radius $x_c$ can be seen in tab. \ref{tab:x_PF}. This expansion leads to broader plasma profiles and decreased core density but are similar in the core and at $x_{PF}$ when magnetically-mapped to the cathode radius $x_c$ as seen in fig. \ref{fig:isat_profile}. Dips between the core ($x/x_c=0)$ and the peak fluctuation region ($x=x_{PF})$ are seen, but fluctuation power from this region ($x/x_c = 0.5$ to $0.7$) is not significant (fig. \ref{fig:isat-fluct-prof}) so this region is not the focus of this study. The line-integrated density as measured by a 56 GHz heterodyne interferometer increases up to $\sim 35\%$ from the M=1 case of $\approx8 \times 10^{13}$ cm$^{-2}$ (fig. \ref{fig:density-line}) but does not increase past a mirror ratio of 2.3.
Discharge power increases only slightly ($3\%$) at higher mirror ratios suggesting negligible impact on density. Langmuir sweeps and triple probe measurements of $T_e$ (\texttt{DR2}) show slightly (less than 25\%) depressed core and slightly elevated edge $T_e$ with increasing mirror ratio (fig. \ref{fig:Te-sweeps}) but otherwise remains unaffected. The temperature affects $I_\text{sat}$ measurements through the $\sqrt{T_e}$ term so small changes are insignificant. The low temperatures indicate that the plasma is collisional given the length scales of the system (as seen in table \ref{tab:plasma-parameters}) and isotropic. Plasma potential decreases across the plasma (fig. \ref{fig:Vp_ExB}) when the mirror ratio exceeds 1.9. This drop in plasma potential may be caused by the grounding of the anode to the wall, which should begin at $M=1.93$ given the $72$ cm anode and $100$ cm vessel diameters. The reason for the local minimum in the M=2.68 is unknown. This potential profile creates a sheared $\boldsymbol{E \times B}$ velocity profile (fig. \ref{fig:Vp_ExB}) limited to $500$ m/s in the core and exceeding $\sim 3$ km/s at the far edge. The flow does not exceed 4\% of the sound speed (tab. \ref{tab:plasma-parameters}) in the core or gradient ($x=x_{PF}$) region. The mirror ratio does not appear to significantly alter azimuthal flow. The floating potential (Vf) profile also exhibits similar behavior to the plasma potential (fig. \ref{fig:Vp_ExB}), but is modified by the presence of primary electrons. 

\subsection{Reduced particle flux}

\begin{figure}
    \centering
    \includegraphics[width=200pt]{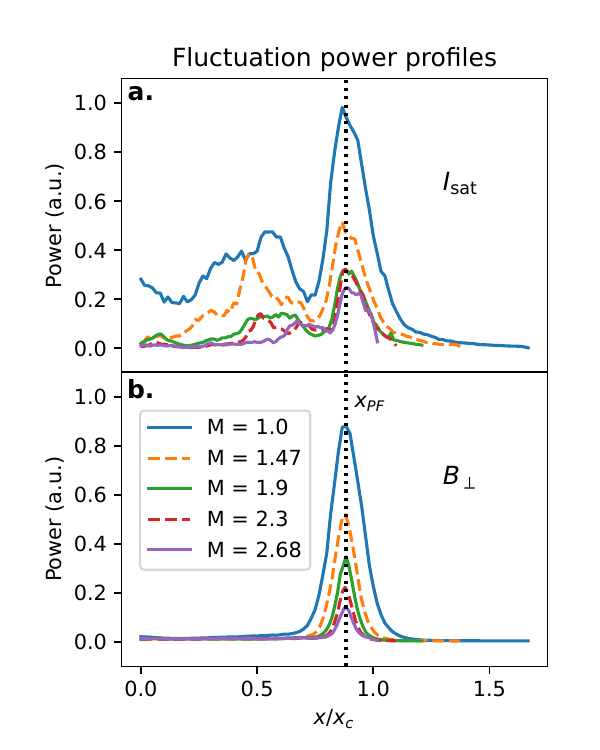}
    \caption{$I_\text{sat}$ (a) and $B_\perp$ (b) fluctuation power profiles for signals 2 kHz and up at z=8.3m (midplane) and z=7.7m, respectively. The lower frequency components in $I_\text{sat}$ are associated with bulk profile evolution, dominate the core region, and are not the focus of this study.}
    \label{fig:isat-fluct-prof}
\end{figure}

The density fluctuation power peaks at the steepest gradient region ($x_{PF} = x/x_c \sim 0.88)$ as expected as seen in fig. \ref{fig:isat-fluct-prof}. $x_{PF}$ occurs at nearly the same magnetically-mapped coordinate for each mirror ratio. These density fluctuations are a large driver of changes in the cross-field particle flux (eq. \ref{eq:flux}). Vf fluctuations also peak at the same location, but the total power across mirror ratios are similar and, relative to density fluctuations, much lower in the core. Core density fluctuations below 2 kHz are substantial in the core at lower mirror ratios, possibly caused by hollow profiles, nonuniform cathode emissivity, or probe perturbations, but are outside the scope of this study. 

\begin{figure}
    \centering
    \includegraphics[width=200pt]{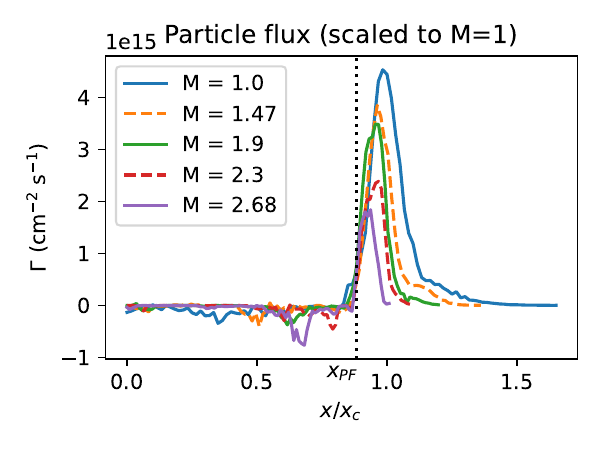}
    \caption{Cross-field, $\tilde{E} \times B$ fluctuation-based particle flux (calculated using eq. \ref{eq:flux}) with respect to mirror ratio. A monotonic decrease in particle flux is observed with increasing mirror ratio at the midplane. Particle flux is normalized by plasma circumference to the $M=1$ case to account for the geometry-induced decrease in particle flux caused by a larger-diameter plasma.}
    \label{fig:particle-flux}
\end{figure}

\begin{figure}
    \centering
    \includegraphics[width=200pt]{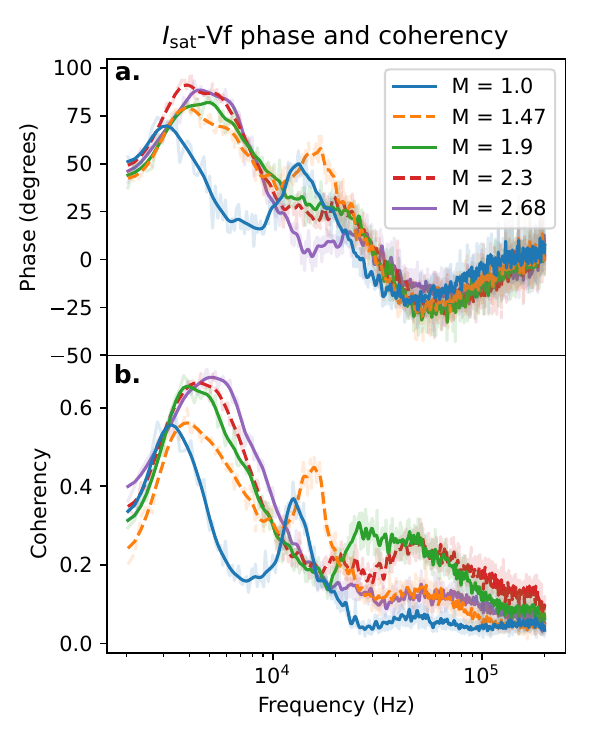}
    \caption{Phase (a) and coherency (b) of $I_\text{sat}$ current and Vf near $x_{PF}$ at the midplane, smoothed. Positive phase means $I_\text{sat}$ leads Vf. Peaks in coherency occur between 3-5 kHz and at the drift-Alv\'en wave peaks between 12 and 25 kHz. These coherency peaks tend to have larger phase shifts than other nearby frequencies. }
    \label{fig:isat-Vf_phase}
\end{figure}

A spectral decomposition technique is used to calculate the time-averaged particle flux \citep{Powers_1974} as seen in fig. \ref{fig:particle-flux}:
\begin{equation}
    \Gamma_{\tilde{E} \times B} = \langle \Tilde{n} \Tilde{v} \rangle = \frac{2}{B} \int^\infty_0 k \left( \omega \right) \gamma_{n \phi} \left( \omega \right) \sin \left( \alpha_{n \phi} \right) \sqrt{ P_{nn} \left( \omega \right) P_{\phi \phi} \left( \omega \right) } d\omega
    \label{eq:flux}
\end{equation}
where $k$ is the azimuthal wavenumber, $\gamma$ is the coherency, $\alpha$ is the cross-phase, and $P$ the power spectrum. This method is more robust than the naive time-integration of $n \left(t \right) \tilde{E} \left( t \right)$ because it accounts for the coherency of the density-potential fluctuations. This representation also enables inspection of each contributing term in the event of surprising or problematic results. A plot of the $I_\text{sat}$-Vf phase can be seen in fig. \ref{fig:isat-Vf_phase}. The flattened particle flux in the core is likely caused by primary electrons emitted by the cathode. These electrons have long mean free paths (greater than a few meters) and sample fluctuations along the length of the machine, mixing the phases of these fluctuations. Since floating potential is set by the hotter electron population, the measured Vf fluctuations are no longer related to the local plasma potential fluctuations of a wave by bulk $T_e$ \citep{Carter_2009}. These primary electrons have a significant effect in the core within the region mapped to the cathode $x \lesssim x_c$. $I_\text{sat}$ fluctuations are not affected.

Azimuthal wave number is measured by two Vf probe tips $0.5$ cm apart. This wavenumber estimation technique yields good agreement with correlation plane measurements (fig. \ref{fig:isat-m-num}). Note that $\tilde{E}$ is not directly measured -- it is instead calculated through the $k(\omega) \sqrt{P_{\phi \phi} (\omega)}$ terms.
The $\tilde{E} \times B$ particle flux clearly decreases with mirror ratio; most of this decrease is attributed to the decrease in density fluctuation power. The particle flux for each mirror ratio was normalized to the $M=1$ case via the plasma circumference to compensate for the increased plasma surface area at the same magnetically-mapped coordinate $x/x_c$. This particle flux is on the order of Bohm diffusion $D_B = \frac{1}{16} \frac{T_e}{B} \approx 6.25 \text{m}^{2} \text{s}^{-1}$ as observed in other transport studies \citep{Maggs-Carter_2008}.

$T_e$ profiles and fluctuations may affect particle fluxes but measurements of both were not taken in the same datarun; nevertheless, a quantification of the effect of $T_e$ on particle flux is attempted.
$T_e$ fluctuations affect $I_\text{sat}$-based density measurements through the $T_e^{-1/2}$ term, and triple probe and Langmuir sweep $T_e$ measurements suggest that temperature gradients have a negligible impact. A naive incorporation of temperature fluctuation data from \texttt{DR2} into particle fluxes from \texttt{DR1} suggest that cross-field particle flux may be underestimated by up to $50\%$ via the $I_\text{sat}$ temperature term, but the trend and relative fluxes across mirror ratios remain unchanged. Such a naive incorporation should be treated with suspicion because of the sensitive nature of the flux with respect to the gradient and the differences in profiles between \texttt{DR1} and \texttt{DR2}. These difference in profiles made be caused by cathode condition, deposits on the anode, or a different gas mix and are difficult to account for.

\subsection{Drift waves}
\begin{figure}
    \centering
    \includegraphics[width=200pt]{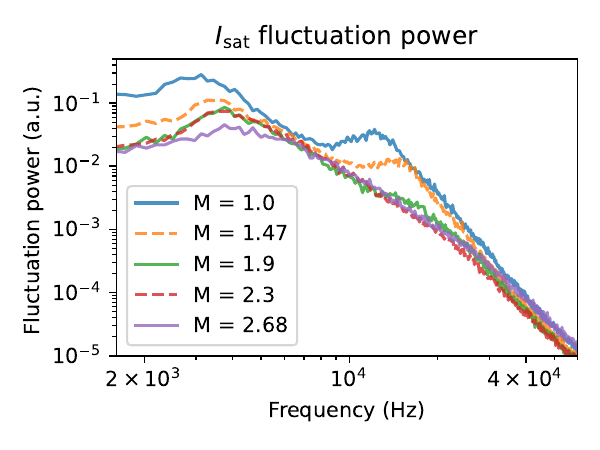}
    \caption{$I_\text{sat}$ (density) fluctuation power averaged over a 1 cm region around $x_\text{PF}$ at the midplane. The fluctuation power is largely featureless below 2 kHz and beyond 40 kHz aside from electronics noise.}
    \label{fig:isat_fluct_power}
\end{figure}

The $I_\text{sat}$ fluctuation power spectra in the region of peak power $x \sim x_\text{PF}$, also where the density gradient is strongest, can be seen in fig. \ref{fig:isat_fluct_power}. Notably, the fluctuation peaks shift to higher frequencies and decrease in total fluctuation power. The shift in frequency may be the Doppler shift caused by the change $\boldsymbol{E \times B}$ plasma rotation seen in fig. \ref{fig:Vp_ExB} at the location $x/x_c \approx x_\text{PF}$. The shift in frequency is somewhat smaller than what would be expected from the field line-averaged increase in Alfv\'en speed at the longest possible wavelength.
The phase angle of $I_\text{sat}$ and Vf provides insight into the nature of the driving instability. Including a nonzero resistivity $\eta$ in the drift wave leads to a small phase shift $\delta$ between density and potential.
This phase shift $\delta$ in a collisional plasmas is on the order of $\delta \approx \omega \nu_e / k_\parallel^2 \bar{v}_e^2$ \citep{Horton_1999}. Estimating this quantity using measured and typical values ($k_\parallel = 0.18$ rad/m, $\bar{v}_e = 1300$ km/s, $\nu_e = 3.7$ MHz, $\omega = 12$ kHz) yields a substantial phase shift of $\delta \approx 46^\circ$, which roughly agrees with the phase shifts in fig. \ref{fig:isat-Vf_phase}, though the implied increased phase shift at higher frequencies does not agree with measurements.
As seen in fig. \ref{fig:isat-Vf_phase}, the phase shift between $I_\text{sat}$ and Vf fluctuations are larger below 10 kHz, implying the presence of additional modes beyond or significant modification of resistive drift wave fluctuations.

\begin{figure}
    \centering
    \includegraphics[width=200pt]{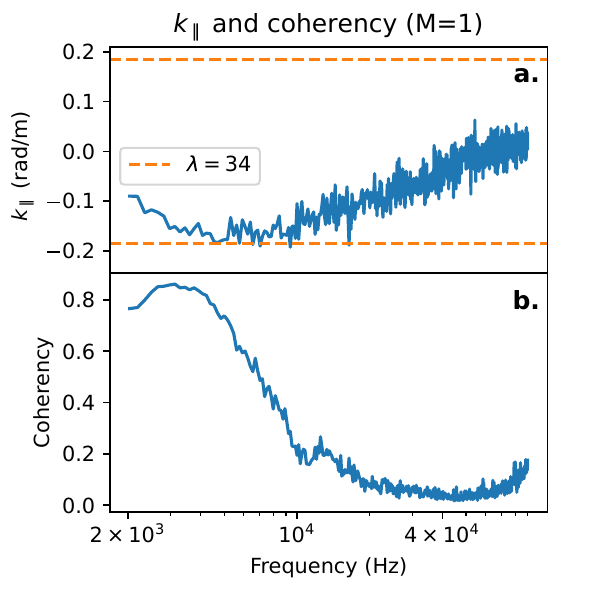}
    \caption{$k_\parallel$ (a) and coherency $\gamma$ (b) as a function of frequency. Only results from the $M=1$ case are available, but it is clear that there are long ($\gtrsim 34$m) wavelength modes at 3 and 12 kHz. The probes used for calculating $k_\parallel$ were located at the midplane (z=8.31) and z=12.14 m, 3.83 m apart.}
    \label{fig:kparallel-coherency}
\end{figure}
The phase difference between two Vf probes, $3.83$ m apart, was used to calculate the parallel wavelength $\frac{2 \pi}{\lambda} = k_\parallel = \phi_{\text{Vf1, Vf2}} / \Delta z$ assuming the wavelengths are greater than $7.66$ m. The two probes mapped to the same field line only in the $M=1$ configuration, so parallel wavenumbers are available only for the flat case. Parallel wavenumbers are theoretically calculable from 2d correlation planes but the coherency dropped dramatically when a mirror geometry was introduced.
A $34$ m wavelength mode likely contributes to the measured $k_\parallel$ from $3$ to  $\gtrsim 10$ kHz (fig \ref{fig:kparallel-coherency}). Drift waves are long-wavelength modes so coherent density and potential fluctuations along the flux tube are expected. The coherency is a measure of similarity of the spectral content of two signals, in this case Vf probes 1 and 2. The coherency is defined as $\gamma = \frac{|\langle P_{1,2} \rangle|}{\langle |P_{1,1}|^2 \rangle \langle |P_{2,2}|^2 \rangle}$ where $P_{x,y}$ is the cross-spectrum between signals $x$ and $y$ and the angle brackets $\langle \rangle$ denote the mean over shots. The coherency between the two Vf probes drops off with increasing frequency, with a slight bump at around 12 kHz. There are several candidates for the driving mechanism of the 3-5 kHz mode, but the 12 kHz mode is most likely a drift-Alfvén wave.

\subsection{\label{sec:sub_turbulence}Turbulence modification}
\begin{figure}
    \centering
    \includegraphics[width=200pt]{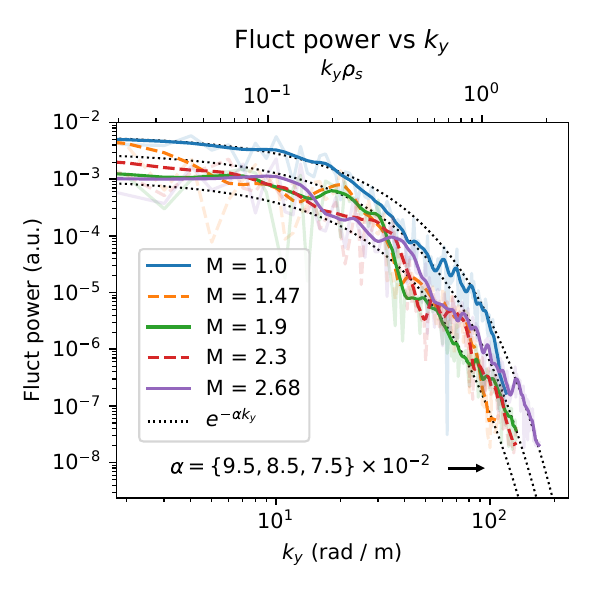}
    \caption{Fluctuation power summed for each $k_y$ for frequencies up to 100 kHz, smoothed. The contribution to fluctuation power is negligible past 100 kHz. The fluctuation power decreases substantially when a mirror configuration is introduced, but no trend is seen otherwise and the $k_y$ spectra remain exponential. Note the logarithmic scale.}
    \label{fig:fluct-power_ky}
\end{figure}

 The wavenumber-power relation in fig. \ref{fig:fluct-power_ky} shows decreased fluctuation power when a mirror configuration is introduced. However, there is no discernible trend when the mirror ratio is increased further. The exponential nature of the curve also remains unchanged. The greatest decrease in fluctuation power occurred in low and high $k_y$'s, around $10$ and $70$ rad/m. The shape of the power-$k_y$ curves follow an exponential distribution, and is inconsistent with a 2d drift-wave turbulent cascade (Wakatani Hasegawa $k^{-3}$) \citep{Hasegawa-Wakatani}. The steep dropoff in fluctuation power with $k_y$ suggests that higher-wavenumber fluctuations do not have a significant effect on transport.

Previous simulations in a flat field \citep{Friedman_simulation_2013} predicted frequency and wavenumber spectra that can be fit with many power laws or exponentials, but the data presented here (figs. \ref{fig:isat_fluct_power}, \ref{fig:bperp_fluct}, \ref{fig:fluct-power_ky}) appear to follow an exponential relationship within measurement variation.

Turbulence measurements can be directly compared to theoretical predictions and other devices, summarized by Liewer \citep{Liewer_measurements_1985}. For saturated drift wave turbulence, one expects the normalized fluctuation level $\tilde{n}/n \sim 1/\langle k_\perp \rangle L_n$, where $k_\perp$ is some typical wavenumber. The power-weighted $k_y$ (calculated from fig. \ref{fig:fluct-power_ky}) was approximately 15 rad/m, which is an order of magnitude too small to satisfy this relationship. $\tilde{n}/n$ scaling with $\rho_s / L_n$, however, is roughly consistent with drift wave turbulence level saturation: the latter is $\approx 3$ times larger. These comparisons suggest that the large, low frequency fluctuations ($\sim 3$ kHz, which had even smaller $k_y$) may have a drift wave turbulence component but are dominantly driven by other instabilities. No trend is seen in $\rho_s / L_n$ and $1/k_y L_n$ when mirror ratio was varied.

Core fluctuations appear to decrease dramatically as seen in the $I_\text{sat}$ fluctuation power (fig. \ref{fig:isat-fluct-prof}). The $I_\text{sat}$ decorrelation time increases from $\sim0.7$ ms for $M=1$ to $\sim2.5$ ms for $M=2.68$. At $x=x_{PF}$, decorrelation times for all mirror ratios remained at $0.2$ ms.

\subsection{Magnetic fluctuations}
\begin{figure}
    \centering
    \includegraphics[width=200pt]{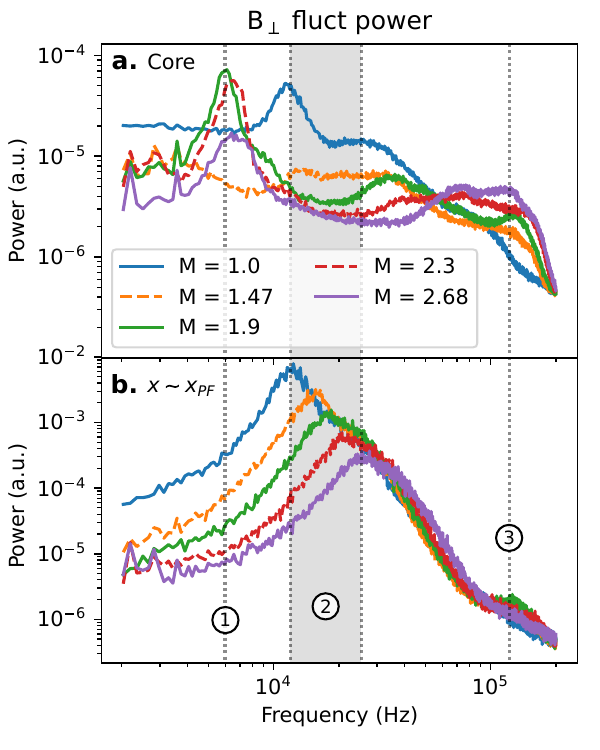}
    \caption{$B_\perp$ fluctuation power averaged at the core from 0 to 3 cm (a) and around the peak fluctuation point $(x \sim x_{PF})$ (b). Fluctuation power decreases across the board with mirror ratio except for core frequencies close to $\Omega_{ci}$. Peaks around $10-30$ kHz at $x_{PF}$ are consistent (region 2) with drift-Alfvén waves and the near-cyclotron frequency features in the core may be resonating Alfvén waves created by the magnetic mirror. Frequencies below 2 kHz and dominated by instrumentation noise and thus excluded.}
    \label{fig:bperp_fluct}
\end{figure}

The perpendicular magnetic fluctuation ($B_\perp$) component of the drift-Alfvén wave can be seen in fig. \ref{fig:bperp_fluct}. These $B_\perp$ fluctuations are spatially and spectrally coincident with the electrostatic fluctuations (fig. \ref{fig:isat_fluct_power}).
Drift-Alfvén waves have been studied in the LAPD in the past \citep{Maggs_1997, Vincena_2006}; strong coupling is observed for $\beta_e > m_e / m_i$ which is satisfied in this study. The Alfvén speed $\omega / k_\parallel = v_A = B/\sqrt{4 \pi n M}$ (given $\omega \ll \Omega_{ci}$) when averaged over the entire column ranges from $\sim 450$ to $\sim 1600$ km/s. A $k_\parallel$ corresponding to a wavelength $\lambda = 34$m roughly falls within the bound established by the kinetic and inertial Alfvén wave dispersion relations at the frequency peaks observed at $x \sim x_{PF}$ seen in fig. \ref{fig:bperp_fluct}. The lengthening of field lines caused by curvature accounts for at most $10\%$ of the change in frequency.

\begin{figure}
    \centering
    \includegraphics[width=200pt]{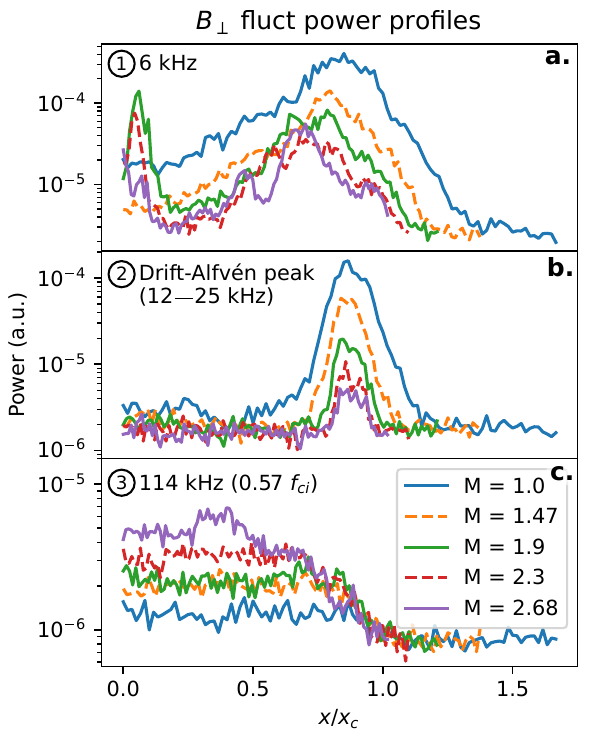}
    \caption{$B_\perp$ fluctuation power profiles for the three regions shown in fig. \ref{fig:bperp_fluct}: region 1 (6 kHz) (a), region 2 — where frequencies are taken from the peaks of the drift-Alfv\'en waves for each mirror ratio (b), and region 3 (114 kHz) (c).}
    \label{fig:Bperp_power_profiles}
\end{figure}

The spatial extent of the $B_\perp$ features identified in fig. \ref{fig:bperp_fluct} are plotted in fig. \ref{fig:Bperp_power_profiles}. Feature 1 at $\approx 6$ kHz shows increased fluctuation amplitudes at $x=0$ for mirror ratios 1.9 and above, but for $M=1$ and $M=1.47$ there is no increase in fluctuation power. A similar feature, but at a much smaller level, is observed in $I_\text{sat}$ fluctuation power in the core as well. This core feature may be caused by the hole in the core seen in the $I_\text{sat}$ profile (fig. \ref{fig:isat_profile}) driving low-amplitude waves or instabilities. Feature 2 in fig. \ref{fig:Bperp_power_profiles} is the magnetic component of the drift-Alfv\'en wave. The fluctuation power peaks at the gradient region and corresponds with the peak in density fluctuations (fig. \ref{fig:isat-fluct-prof}). 

\begin{figure}
    \centering
    \includegraphics[width=200pt]{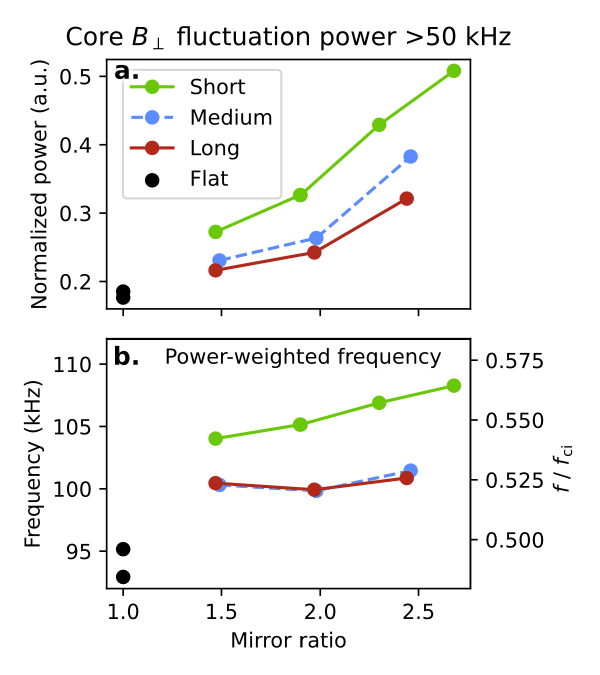}
    \caption{Summed fluctuation power of $B_\perp$ in the core ($x/x_c \leq 0.3$) as a function of mirror length and ratio. Top (a): the fluctuation power is normalized by the sum of the full-spectrum summed power. Bottom(b): the frequency of the power distribution > 50 kHz weighted by the fluctuation power.}
    \label{fig:Bperp_core_highfreq}
\end{figure}

Feature 3 is particularly interesting because
this the only fluctuating quantity to \textit{increase} with mirror ratio, seen in fig. \ref{fig:Bperp_core_highfreq}. This feature may be broad evanescent Alfv\'enic fluctuations from the plasma source.
These fluctuations have been observed in the LAPD in the source region alongside an Alfvén wave maser \citep{Maggs_2005}. Note that the Alfv\'en maser cannot enter the mirror cell at mirror ratios greater than 1.75 because the Alfv\'en maser resonates at 0.57 $f_{ci}$ but the midplane is always at or near 500G.

The sub-$2$ kHz modes in $B_\perp$ and its harmonics are nearly constant in power across the entire plasma; these features are likely perturbations from the magnet power supplies and thus ignored. The lack of radial, azimuthal, and axial structure in these magnetic signals below 2 kHz and narrow bandwidth indicate a non-plasma origin. Significant radial and azimuthal structure in $B_\perp$ fluctuation power starts to appear in frequencies larger than 4 kHz.

\section{\label{sec:structure}2d Structure}
\begin{figure}
    \centering
    \includegraphics[width=200pt]{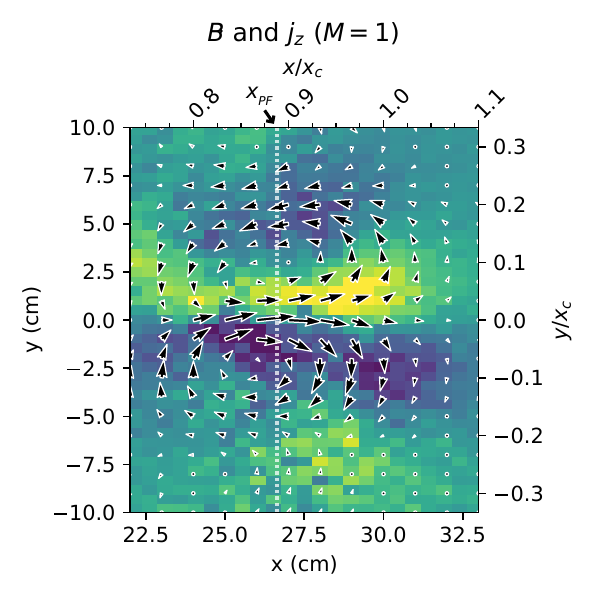}
    \caption{Perpendicular magnetic field and the derived current density for the flat-field ($M=1$) case using a Bdot probe with an axially-separated $I_\text{sat}$ reference (\texttt{DR2}). The x-y plane was centered near $x_{PF}$.}
    \label{fig:Bvec_jz_M=1}
\end{figure}

The perpendicular magnetic field structure is measured by collecting x-y planar bdot ($d B_{\{x, y, z \} } / dt$) data alongside a stationary, axially separated $I_\text{sat}$ reference probe (\texttt{DR2}). This probe provides a phase reference for the magnetic field fluctuations, allowing a 2d map of relative phase to be constructed over many shots. Only the region around $x_{PF}$ was measured because of constraints on probe movement. The amplitude and phases for each magnetic field component are then used to reconstruct the local magnetic fluctuation vector $\boldsymbol{B}$. The axial current density structure, $j_z$, can be derived from this vector field. $\boldsymbol{B}$ and the corresponding $j_z$ for the flat-field ($M=1$) case can be seen in fig. \ref{fig:Bvec_jz_M=1}.
Two main current channels can be seen with the magnetic fields circulating around them. This structure quickly decoheres in time as expected in a turbulent plasma. At higher mirror ratios, the field magnitude and corresponding current density decrease (which was also seen in \texttt{DR1}: fig. \ref{fig:bperp_fluct}).

\begin{figure}
    \centering
    \includegraphics[width=200pt]{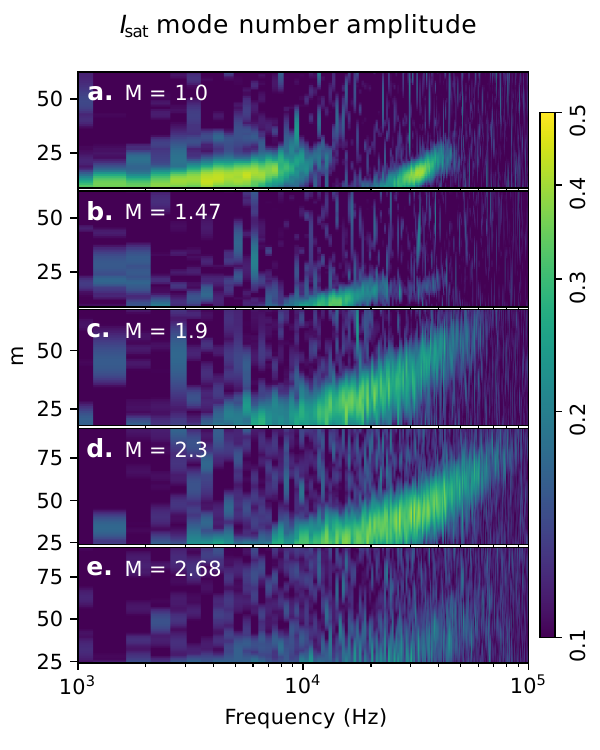}
    \caption{Azimuthal mode number $m$ amplitudes calculated from two axially-separated, correlated, $I_\text{sat}$ probes. Increasing mirror ratio (a to e) leads to increased $m$ at higher frequencies. (\texttt{DR2})}
    \label{fig:isat-m-num}
\end{figure}

\begin{figure}
    \centering
    \includegraphics[width=200pt]{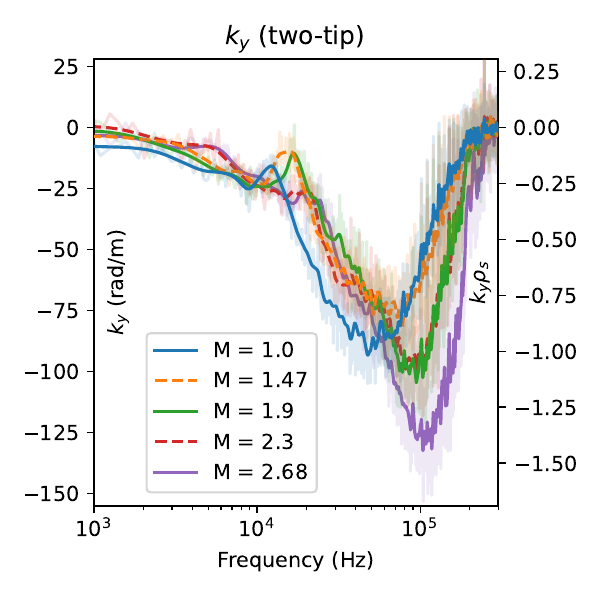}
    \caption{$k_y$ averaged about $x_{PF}$ and smoothed for each mirror ratio calculated using two vertically-separated Vf tips on the same probe. Little change is seen in $k_y$ at lower frequencies but higher frequencies tend towards larger $k_y$ at higher mirror ratios.}
    \label{fig:kperp}
\end{figure}

Using two, axially-separated, correlated $I_\text{sat}$ measurements (\texttt{DR2}), with one collecting x-y planar data, the azimuthal mode number $m$ (radially integrated) was calculated. Higher-frequency and higher-$m$ features are seen with increasing mirror ratio (fig. \ref{fig:isat-m-num}). The increased frequencies may be caused by a  change in Doppler shift by the $\boldsymbol{E \times B}$ flow. This higher-m trend suggests that azimuthal structures do not scale with increased plasma radius but instead remain roughly the same size. The limited planar probe movement caused an increase in the lower bound on $m$  in higher mirror ratios. At mirror ratios 1.47 and higher, the lower frequency component ($< 10$ kHz) appears to decrease significantly in amplitude. 
Calculating $k_\perp$ from $m$ evaluated at $x \sim x_c$ yields similar $k_y$ values as the two-tip technique (fig. \ref{fig:kperp}). The average $k_y$ for a given frequency can be calculated using two Vf tips on the same probe by calculating the phase difference and dividing by the spatial separation of $5$ mm: $k_y = \phi_{\text{vf1, vf2}} / \Delta y$ \citep{Powers-twoprobe-ky}. The maximum $|k_y|$ measurable before aliasing is $\pi / \Delta y \approx 628$ rad/m. As seen in fig. \ref{fig:kperp}, the $k_y$ spectrum remains similar across mirror ratios, but the wavenumber extends further into higher frequencies with increasing mirror ratio. 
These azimuthal mode numbers and gradient scale lengths are consistent with linear simulations using the 3d fluid code \texttt{BOUT} \citep{Popovich_2010} in the flat, unbiased case.

\section{\label{sec:discussion}Discussion}
\subsection{Lack of mirror-driven instabilities}
No evidence is seen for mirror-driven instabilities — curvature, loss-cone, or otherwise. Given the LAPD parameters in this study (tables \ref{tab:fields} and \ref{tab:plasma-parameters}), the collision frequencies are sufficiently high such that the mirror is in the gas-dynamic regime: losses out of the mirror throat are governed by gas-dynamic equations rather than free streaming through the loss cone. To be in the gas-dynamic regime, the mirror length must exceed the mean free path of the ions \citep{Ivanov_2013}:
\begin{equation}
    L > \lambda_{ii} \ln{M} / M
\end{equation}
where $L$ is the mirror length, $\lambda_{ii}$ is the ion mean free path, and $M$ is the mirror ratio.
These collisions populate the loss cone and maintain a (cold) Maxwellian distribution, eliminating the possibility of loss-cone-, ion-driven instabilities like the AIC \citep{Casper_1982} or DCLC \citep{Simonen_1976,Kanaev_1979} instabilities that have been observed in other (historic) devices. 

The paraxial, approximate interchange growth rate is \citep{Post_1987,Ryutov_2011}
\begin{equation}
    \Gamma_0 = \frac{c_s}{\sqrt{L_M L_P}}
\end{equation}
which yields $\Gamma_0 \approx 1.2$ kHz using $L_M \approx 7$m and $L_P = 17$m. $c_s$ is used instead of $\bar{v_i}$ because $T_i \ll T_e$ and mirror length $L$ is split to distinguish between the contributions of the plasma length and mirror length to inertia and to curvature drive, respectively.
Interchange is not visible in-part because the aspect ratio of these mirrors is quite large, limiting the growth rate of interchange. The length of the mirror (3.5 m), radius of curvature (6-7 m), and plasma column (17 m) are much larger than the radius of the plasma (0.5 m maximum), so the plasma inertia is large relative to the instability drivers. Line-tying to the cathode may further lower the growth rate. The hot cathode used for plasma formation could function as a thermionic endplate that can supply current to short out the flute-like interchange perturbations. Line-tying has been seen in flux rope experiments on the LAPD using a hotter, denser source \citep{Compernolle_2011}, also in other devices \citep{Fornaca_1979}, and is why interchange was not seen in the earliest mirror machines \citep{Post_1987}. Note that the plasma terminates on the cathode or end plates before the magnetic field flares out, so there is no contribution to stability from an expander tank as seen in other GDTs \citep{Ryutov_2011,Ivanov_2013}. 
Finite Larmor radius (FLR) effects may provide a stabilizing effect for larger azimuthal mode numbers. 
At the highest mirror ratio, assuming a plasma radius of $a_0 =\sqrt{2.68} * x_{PF}=43$ cm, the FLR stability criterion $\frac{m}{2} \frac{\rho_i L}{a_0^2} > 1$ \citep{Ryutov_2011} suggests a stabilizing effect may be present for azimuthal mode numbers $m > 4$.

If the curvature-induced interchange instability were observable, then introducing a mirror configuration would lead to new features in $I_\text{sat}$ and Bdot fluctuations. In particular, low-frequency mode(s)  -- likely less than 10 kHz given the low m-number and plasma rotation rates -- would be observed growing from the pressure gradient region. For onset of the interchange instability, the mirror curvature or plasma pressure would need to be increased but the precise conditions required for this onset are not yet known for the LAPD.

Interchange could also be at least partially stabilized by the continuous production of electrons in the core that are electrostatically trapped by the ambipolar potential \citep{Guest_1971}. The intuition behind this stabilization mechanism is as follows: electrons are continuously produced via ionization of neutrals, and any change in the local potential will cause more or fewer electrons to be lost out the ends of the device along that field line, counteracting the potential change. This stabilization mechanism has been experimentally demonstrated  to completely suppress interchange when the ambipolar potential $\Phi \gtrsim 6T_e$ \citep{Komori_1987}.

The $\boldsymbol{E \times B}$ shear flow present (fig. \ref{fig:Vp_ExB}) may also make a contribution to the stabilization of interchange \citep{Ryutov_2011,Bagryansky_2003,Bagryansky_2007,Beklemishev_2010}. The estimated shearing rate is between 3 and 10 kHz, which is greater than the estimated $\approx 1.2$ kHz growth rate of the interchange mode.

\subsection{Instabilities driving turbulence}

Rotational interchange can be significant driver of the broadband turbulence spectrum in the LAPD, particularly when a biased limiter is installed. This observation has been confirmed by both linear simulations \citep{Popovich_2010} and biasing experiments \citep{Schaffner_2013}. 

This rotational interchange mode has the following attributes, as summarized by \citet{Jassby_transverse_1972}: flute-like ($k_\parallel=0)$, $|e\tilde{\phi}/T_e|/|\tilde{n}/n| \gtrsim 1$, radial potential phase variation $45$ to $90^\circ$, maximum possible $|e\tilde{\phi}/T_e| < 1$. All of these attributes are seen for the lower frequency (3 kHz) mode. The Vf radial phase variation when $M>1$ is not clearly seen because the coherency is dramatically reduced along the field line.
The rotational interchange mode could couple with the drift wave at $k_\parallel = \pi / L \sim 0.37$ rad/m ($n=0.5$), which has been observed in the past \citep{Schaffner_2013} and likely present here.
Estimates of shearing rate from the $\boldsymbol{E \times B}$ flow velocity profile (fig. \ref{fig:Vp_ExB}), calculated fluctuation ratios, and radial phase shift variation suggest that Kelvin-Helmholtz-driven turbulence is not significant, if present at all. Historically, biasing a limiter has been required to clearly observe the Kelvin-Helmholtz instability \citep{horton_vorticity_2005, Schaffner_2012, Schaffner_2013}.

Low frequency density fluctuations may also be driven by a flute-like conducting-wall temperature-gradient instability which only requires an electron temperature gradient to grow (even with straight field lines) \citep{Berk_1991}. Simulations of turbulence in the LAPD suggest the possible presence of these conducting wall modes (CWM) which have the highest growth rate for $m \leq 20$ \citep{Friedman_2013}. This lower-$m$ mode could be responsible for the peak around $3$ kHz in the $M=1$ $I_\text{sat}$ fluctuation (fig. \ref{fig:isat_fluct_power}) and azimuthal mode numbers (fig. \ref{fig:isat-m-num}) and for the low-frequency low-$k_\parallel$ or flute-like behavior (fig. \ref{fig:kparallel-coherency}). This CWM may also be responsible for flatter electron temperature profiles seen in previous studies \citep{Perks_impact_2022, Schaffner_2013} (fig. \ref{fig:Te-sweeps}). 

These linearly unstable modes may be outgrown by a rapidly-growing nonlinear instability that couples to drift-like modes as suggested by simulations \citep{Friedman_2013}. This nonlinear instability is driven by the density gradient at an axial modenumber of $n=0$ and nonlinearly transfers energy to $n \neq 0$ fluctuations. 

The conducting wall mode and nonlinear instability may be present in these mirror experiments but the spectra are adequately explained by linearly unstable modes. Precise identification these modes requires further study; neither of these instabilities have been directly observed in the LAPD.

\subsection{Causes of particle flux reduction}

\begin{figure}
    \centering
    \includegraphics[width=200pt]{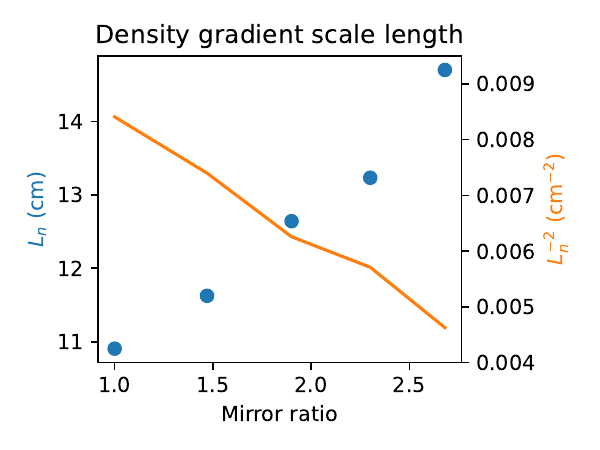}
    \caption{Gradient scale length $L_n$ and the associated term in the drift wave growth rate $L_n^{-2}$. This scale length was calculated over a 3 cm region around $x_\text{PF}$ (peak fluctuation region) at the midplane. Increasing the mirror ratio increases the gradient scale length, which suggests weakening of the underlying instability driver.}
    \label{fig:L_n}
\end{figure}

The reduction in particle flux explained by a reduction in density fluctuations likely caused by a increased gradient scale length $L_n = \frac{n}{\nabla n }$ (fig. \ref{fig:L_n}), decreasing the linear drift wave growth rate and saturation level seen in sec. \ref{sec:sub_turbulence}. This gradient length reduction may also reduce the growth rate of the rotational interchange instability, which may be the dominate driver for the low-frequency large-amplitude density fluctuations.

The decorrelation time of $I_\text{sat}$ time series data is around 0.15 ms at $x_\text{PF}$. An estimate of the $\boldsymbol{E \times B}$ flow shear from fig. \ref{fig:Vp_ExB} (\texttt{DR2}) yields a shearing time between 0.1 and 0.3 ms at $x_\text{PF}$. These times suggest that spontaneous flow shear may be important for suppressing turbulence, as seen in other studies \citep{Schaffner_turbulence_2013,Cho_2005}, at all mirror ratios. However, no clear trend in shearing strength is seen with mirror ratio.

It is important to note that the electron thermal diffusion time along the field line is very long compared to the frequency of the drift wave ($\omega \gtrsim k_\parallel \bar{v}_e^2 / \nu_{ei}$) \citep{Goldston_textbook} so the electron temperature along the field line may not be constant on the drift wave timescale. This factor is not taken into account in this analysis but may have substantial impact on interpretations of the measured phase shift.

\section{\label{sec:conclusion}Conclusions and future work}

Turbulence and transport was studied in mirrors with varying lengths and ratios using the flexible magnetic geometry of the LAPD. Particle flux and fluctuation amplitudes decreased up to a factor of two when mirror ratio was increased. The primary drivers of turbulence were identified as the rotational interchange mode, caused by spontaneous rotation, and unstable drift-Alfvén waves driven by the density gradient. The decrease in density fluctuation amplitudes can be attributed to an increase in the gradient scale length caused by the dimensionally wider plasma at the mirror midplane. Despite imposing a mirror configuration, no signs of mirror-driven instabilities were observed. The highly-collisional, GDT-like plasma produced suppressed any velocity space instabilities. The interchange growth rate was likely suppressed to an undetectable level by line-tying, in-cell electron production, and shear flow.

Future experiments in hotter regimes with the new LaB6 cathode \citep{LAPD_LaB6} will need to be performed to evaluate the robustness of these results, particularly concerning the stabilization of curvature-induced interchange. Additionally, the source field should be matched to the mirror midplane field so that the plasma remains the same radius to isolate geometric effects.
Simultaneous measurements using flux and/or vorticity probes and $I_\text{sat}$ are needed to concretely determine if azimuthal flow shear is modified by the mirror field, and to quantify the effect of flows on rotational interchange and drift wave instability drive in general.
Multiple simultaneous axial measurements of potential would enable better understanding of the axial wavenumber and identification of possible modes.

This work was performed at the Basic Plasma Science Facility, which is a DOE Office of Science, FES collaborative user facility and is funded by DOE (DE-FC02-07ER54918) and the National Science Foundation (NSF-PHY 1036140). The authors would like to thank Dr. Giovanni Rossi and Dr. Steven Vincena for probe setup and assistance with data collection, and Dr. Mel Abler, Prof. Walter Gekelman, and Prof. George Morales for insightful discussion.

The data and code supporting this study are available from the corresponding author upon reasonable request.

Competing interests: The author(s) declare none.

\section{Bibliography}
\bibliographystyle{jpp}
\bibliography{_bibliography}%

\end{document}